# A Novel Approach for Clone Group Mapping by using Topic Modeling

Ruixia Zhang, Liping Zhang, Huan Wang and Zhuo Chen

Computer and information engineering college, Inner Mongolia normal university, Hohhot, China
zhangruixia923@163.com

**ABSTRACT.**

*Clone group mapping has a very important significance in the evolution of code clone. The topic modeling techniques were applied into code clone firstly and a new clone group mapping method was proposed. The method is very effective for not only Type-1 and Type-2 clone but also Type-3 clone .By making full use of the source text and structure information, topic modeling techniques transform the mapping problem of high-dimensional code space into a low-dimensional topic space, the goal of clone group mapping was indirectly reached by mapping clone group topics. Experiments on four open source software show that the recall and precision are up to 0.99, thus the method can effectively and accurately reach the goal of clone group mapping.*

## 1. INTRODUCTION

A code clone is a code portion in source files that is identical or similar to another[1]. It is suspected that many large systems contain approximately 9%-17% clone code, sometimes as high as even 50%[2]. The activities of the programmers including copy, paste and modify result in lots of code clone in the software systems.

In the past and in recent years, several researchers have indicated a number of factors that affect source code maintainability. So Code clones have been considered as a bad software development practice. For example, when a cloned code fragment needs to be changed, for example because of a bug fix, it might be necessary to propagate such a change across all clones. Software maintenance and evolution are crucial activities in the software development lifecycle, impacting up to 80% of the overall cost and effort [3].Although there is a common understanding that cloning is a bad practice, recent studies have shown that clones are not necessarily a bad thing. For example, copying source code without defect can reduce the potential risk of writing new code, save development time and cost. Kapser and Godfrey[4] said, in many cases cloning has been used as a development practice, and developers are often able to handle "harmful" situations.

To avoid problems clones can cause due to change mis-alignment and exploit the advantages of clones, it is necessary to provide the developers with method able to support clone tracking. Therefore，in order to meet the demands of clone evolution, an clone mapping method is put forward.The topic modeling techniques is applied to code clone firstly and a new clone group mapping method is proposed.

## 2. TERMS AND DEFINITIONS

In this section, we introduce definition of code clone and relevant terminology.

A **code clone** is a code fragment (CF) in source files that is identical or similar to another. A clone fragment is a contiguous section of source code. As such, a clone fragment has a well-defined start and end and is contained in exactly one file in exactly one version of the system. A **clone pair** relates two fragments that are similar to each other and are contained in the same version of the system. In contrast to a clone pair, a **clone group** is composed of two or more similar fragments. **Clone group mapping** reflect how an clone group evolve from a previous version to the current version, is the core technology in the evolution of code clone across versions.

According to the degree of similarity between clones, clone is divided into four different types:

Type-1Clone(CF and CF1)：Identical code fragments except for variations in white-spaces and comments.

Type-2 Clone(CF and CF2)：Structurally/syntactically identical fragments except for variations in the names of identifiers, literals, types, layout and comments.

Type-3 Clone(CF and CF3)：Code fragments that exhibit similarity as of Type-2 clones and also allow further differences such as additions, deletions or modifications of statements.

Type-4 Clone(CF and CF4)：Code fragments that exhibit identical functional behaviour are implemented through very different syntactic structures.

Figure 1 shows the above four kinds of clones. Type-1 Clone is often called exact clone .Type-2 and type-3 clones are often collectively called near-miss clone.

Figure 1. Four kinds of clones

## 3. RELATED WORK

Software development and maintenance in practice follow a dynamic process. With the growth of the program source, code clones also experience evolution from version to version. What change the Clone group have happened from one version to next need to be made judgments by clone group mapping.

To map clone group across consecutive versions of a program, mainly five different approaches have been found in the literature.

**Based on text[5]**：It separates clone detection from each version, and then similarity based heuristic mapping of clones in pairs of subsequent versions. Text similarity are often computed by the Longest Common Subsequence(LCS)or Edit Distance(Levenshtein Distance, LD) algorithm that have quadratic runtime, which lead to inefficient clone mapping. The method is susceptible to large change in clone.

**Based on version management tools (CVS or SVN)[6]**：Clones detected from the first version are mapped to consecutive versions based on change logs obtained from source code repositories. It is faster than the above technique, but can miss the clones introduced after the first version.

**Based on incremental clone detection algorithm[7]**：Clones are mapped during the incremental clone detection that used source code changes between revisions. It can reduce the redundant computation and save time .So it is faster than the above two techniques, but cannot operate on the clone detection results obtained from traditional non-incremental tools.

**Based on functions[8]:** It separates clone detection from each version, functions are mapped across subsequent versions, then clones are mapped based on the mapped functions. To some extent, it improves the efficiency and accuracy of the mapping, but it is susceptible to similar overloaded/overridden functions for its over-reliance on function information.

**Based on CRD(Clone Region Descriptor)[9]**：Clone code is represented by CRD, then clones are mapped based on CRD between versions. It is not easily influenced by position of the code clone. Mutations or big difference between versions can reduce the mapping validity greatly.

## 4. APPROACH

In this section we present a new clone group mapping approach based on topic modeling for tracking clone groups across different versions.

### 4.1 Overview of Topic Model

Topic models are generative probabilistic models, originally used in the area of natural language processing for representing text documents. LDA (Latent Dirichlet Allocation) has recently been applied to a variety of domains, due to its attractive features. First, LDA enables a low-dimensional representation of text, which (i) uncovers latent semantic relationships and (ii) allows faster analysis on text [10]. Second, LDA is unsupervised, meaning no labeled training data is required for it to automatically discover topics in a corpus. And finally, LDA has proven to be fast and scalable to millions of documents or more [11]. For these reasons, in this paper we use LDA as our topic model.

In the LDA model, LDA is statistical models that infer latent topics to describe a corpus of text documents [12]. Topics are collections of words that co-occur frequently in the corpus. For example, a topic discovered from a newspaper collection might contain the words {cash bank money finance loan}, representing the "finance industry" concept; another might contain {fish river stream water bank}, representing the "river" concept. So, documents can be represented by the topics within them. Topic modeling techniques transform the text into topic space.

Recently, researchers found topics to be effective tools for structuring various software artifacts, such as source code, requirements documents, and bug reports. Kuhn [13] made the first attempt to apply topic modeling technique to source code, and tried to discover the functional topics. W. Thomas[14]performed a detailed investigation of the usefulness of topic evolution models for analyzing software evolution, they found that topic models were an effective technique for automatically discovering and summarizing software change activities. Asuncion[15] used the topic modeling techniques to study software traceability. In addition, topic model was also used to study class cohesion [16]and bug location[17].

### 4.2 Mapping Clone Group Based on Topic Modeling

**4.2.1 Framework of The Algorithm.**

The paper uses the LDA topic modeling technology to map clone group. It mainly works in the following three steps: (1) extracting the topics from clone group, (2) calculating the similarity between topics,(3) mapping clone group topics. Figure 2 shows the framework of the algorithm. Let $CG^n = \{cg_1^n, cg_2^n, \cdots cg_s^n\}$ be the reported clone groups in $V_n$, $T^n = \{t_1^n, t_2^n, \cdots t_s^n\}$ refers to the clone group topics extracted from the clone group $CG^n$ where $t_i^n$ was extracted from $cg_i^n$, $1 \leq i \leq n$.

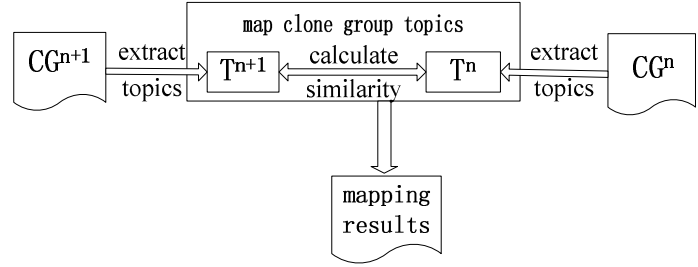

Figure 2. The frame of mapping algorithm

To track clone groups over two different versions $V_{n+1}$ and $V_n$, we compare every clone group in version $V_{n+1}$ to every clone group in version $V_n$. Topic modeling technology is used to extract the topics from each clone group in the version $V_n$ and $V_{n+1}$ respectively. At this point, since topic is the only representation of corresponding clone group, the problem of mapping clone group between two versions of a program is reduced to the mapping of clone group topics between two versions. Then clone group topics are mapped by comparing similarity between topics in the version $V_{n+1}$ and $V_n$. If the topic $t_i^{n+1}$ of a clone group $cg_i^{n+1}$ in version $v_{n+1}$ matches to the topic $t_j^n$ of a clone group $cg_j^n$ in the version $V_n$, we know that the clone group $cg_i^{n+1}$ in $V_{n+1}$ and the clone group $cg_j^n$ in $V_n$ are the same. Due to the transitivity of the relation of equivalence, we can then conclude that clone group $cg_i^{n+1}$ is related to clone group $cg_j^n$. The algorithm is as follows:

**Clone Group Mapping algorithm**
- ∀ $cg_i^{n+1} \in CG^{n+1}$, $CG^{n+1}$ in $V_{n+1}$
  - extract $t_i^{n+1}$ from $cg_i^{n+1}$
  - ∀ $cg_j^n \in CG^n$, $CG^n$ in $V_n$
    - extract $t_j^n$ from $cg_j^n$
    - *caculate similarity between $t_j^n$ and $t_i^{n+1}$, and store similarity value in the array unit sim [j]*
  - *suppose sim[k]=max{sim[ ]}:*
    IF $sim[k] \geqslant \delta$, $cg_i^{n+1}$ is mapped back to $cg_j^n$, namely $cg_i^{n+1} \longrightarrow cg_j^n$ ;
    THEN $cg_i^{n+1} \longrightarrow null$.
- *Return all mapping results*

**4.2.2 Extract Clone Group Topics**

Under the standard programming style, software is suitable for extracting the topics using the LDA model. The paper uses MALLET topic modeling toolkit to extract the topics. MALLET is a Java-based package for statistical natural language processing, document classification, clustering, topic modeling, information extraction, and other machine learning applications to text. It contains efficient, sampling-based implementations of LDA. Figure 3 shows the topics extracted from source code of clone group by MALLET.

**Preprocess the source code.** Each topic is collections of words that co-occur frequently in the clone group, and is the only representation of corresponding clone group. The topic contains a large number of irrelevant information, such as stop words,comments, which play a small role in characterization of clone group information. So, before extracting the topics, we remove irrelevant information, mainly：1）comments of the source code. 2）programming language keywords, such as "for", "return", and "class", etc. 3) Programming related words, such as "main", "arg", and "util", etc. 4）common English language stop words, such as "the", "it", and "on", etc.

**Choose the number of topic.** The proper number of topic is a key to influence the accuracy of clone group mapping. For any given corpus, there is no provably optimal choice for the number of topics. The choice is a trade-off between coarser topics and finer-grained topics. setting the number of topics to extremely small values results in topics that contain multiple concepts, while setting the number of topics to extremely large values results in topics that are too fine to be meaningful and only reveal the idiosyncrasies of the clone group.

In the paper, through experimental analysis, it is best for setting the number of topics to one. In the same clone group, clone code is a code portion that is identical or similar to another. The whole clone group is multiple copies of the same clone whose syntactic or semantic function is same. In other words, a clone group can be represented by a topic.

```
- <topics>
    - <topic titles="tmplist, false, tmpdoc, tdocument, bfwin, save, backend, modified,
        data, documentlist" totalTokens="62" alpha="-1.001" id="0">
        <word count="12" weight="0.1935483870967742">tmplist</word>
        <word count="8" weight="0.12903225806451613">false</word>
        <word count="8" weight="0.12903225806451613">tmpdoc</word>
        <word count="4" weight="0.06451612903225806">list</word>
        <word count="4" weight="0.06451612903225806">tdocument</word>
        <word count="4" weight="0.06451612903225806">bfwin</word>
        <word count="4" weight="0.06451612903225806">save</word>
        <word count="2" weight="0.03225806451612903">backend</word>
        <word count="2" weight="0.03225806451612903">doc</word>
        <word count="2" weight="0.03225806451612903">modified</word>
        <word count="2" weight="0.03225806451612903">data</word>
        <word count="2" weight="0.03225806451612903">documentlist</word>
        <word count="2" weight="0.03225806451612903">glist</word>
        <word count="2" weight="0.03225806451612903">tbfwin</word>
        <word count="1" weight="0.016129032258064516">widget</word>
        <word count="1" weight="0.016129032258064516">gtkwidget</word>
        <word count="1" weight="0.016129032258064516">cb</word>
        <word count="1" weight="0.016129032258064516">file</word>
    </topic>
</topics>
```

Figure 3.The topics extracted by MALLET

### 4.2.3 Mapping Clone Group Topics

A clone group has an arbitrary number of ancestors which are its occurrences in the previous version of the system. Each clone group has zero or one descendant which is its occurrence in the next version of the system. Clone group mapping is determined by the degree of similarity between clone groups in the different versions.

**The threshold of clone group mapping** . Clone group mapping is determined indirectly by similarity between clone group topics from different versions. If the similarity between the topic $t_j^n$ and topic $t_i^{n+1}$ is highest, and the similarity values is not less than certain threshold ($\delta$).In that way, we can conclude that the topic $t_i^{n+1}$ is mapped back to the topic $t_j^n$, namely the clone group $cg_i^{n+1}$ is mapped back to the clone group $cg_j^n$ . In the paper, Similarity threshold $\delta$ is set to 0.8. That's because the similarity value between $t_i^{n+1}$ and $t_j^n$ vary from 0.8 to 1 when $cg_i^{n+1}$ is mapped back to the clone group $cg_j^n$, and the similarity value between $t_i^{n+1}$ and $t_j^n$ is less than 0.8 when $cg_j^n$ is not origin of clone group $cg_i^{n+1}$. So, setting similarity threshold $\delta$ is 0.8.

**Map to origin clone group**. The number of clone group is generally on the rise in the process of software evolution. If the mapping is carried out from the version $V_n$ to $V_{n+1}$, new clone groups are failed to map. On the contrary, disappeared clone groups are failed to map. However, we are more interested in clone code near to the current version in the study of clone evolution. That is to say, compared with disappeared clone group, we are more interested in new clone group. So, the mapping is carried out from the version $V_{n+1}$ to $V_n$.

## 5. CASE STUDY

### 5.1 Systems Under Study

Due to the difference in size of software system, number of clone group in each version ranging from dozens to thousands ,in view of the limitations of manually inspection, so we perform case study on the source code of four small and medium-sized, open source software systems which is written in different programming languages. The detail of software is shown in Table 1. NiCad is used to detect clone code. NiCad, a clone detector, can detect Type-1 、Type-2 and Type-3 clones written in multiple programming language（C、JAVA、C#）and have a high precision rate and recall rate.

Table 1. The detail of software

| software | Bluefish | MALLET | ArgoUML | PostgreSQL |
|---|---|---|---|---|
| Implementation language | C | JAVA | JAVA | C |
| Average size | 23MB | 31MB | 35MB | 92MB |
| Number of the selected version | 2 | 3 | 3 | 4 |
| Number of Clone group (on average) | 20 | 145 | 299 | 506 |

### 5.2 Evaluation Measures

To evaluate the feasibility and validity of the approach, we use Precision and Recall as Evaluation Measures to manually inspect the results of the approach based on topic modeling . Precision and Recall are defined as follows:

**Recall:** Of all the actual clone group mappings, how many were discovered?

$$\text{Recall} = \frac{\text{the number of correct mapping}}{\text{the number of actual clone group mappings}}$$

**Precision:** Of all the clone group mappings discovered, how many are correct?

$$\text{Precision} = \frac{\text{the number of correct mapping}}{\text{the number of correct and incorrect mapping}}$$

### 5.3 Results

Clone group mapping is carried on consecutive versions of other software, and manually inspect the Precision and Recall of the mapping results. The results can be seen in Table 2 and Table 3. The Precision and Recall of the approach are as high as 0.99, which reveal the validity and feasibility of clone group mapping approach based on topic modeling. The runtime of clone group mapping across versions is acceptable. But the results are enough to reveal the feasibility of the approach. This method can effectively achieve the clone group mapping.

Unlike the mapping method based on text, the basic collection of the mapping is the clone group topics, not intermediate representation of clone code(e.g., token and AST). Topic has a large granularity and the higher level of abstraction. However, the difference of topics between different clone groups in the same version is very large and the difference of topics between same clone groups in the different version is very little, which make the clone group mapping method based on topic modeling practicable and effective.

Table 2. The experimental results of the approach

| Software and versions<br>Evaluation Measures | Bluefish | | PostgreSQL | | PostgreSQL | | PostgreSQL | |
|---|---|---|---|---|---|---|---|---|
| | 2.2.4 | 2.2.3 | 9.1.5 | 9.1.4 | 9.1.4 | 9.1.3 | 9.1.3 | 9.1.2 |
| Precision | 1 | | 0.996 | | 0.996 | | 0.994 | |
| Recall | 0.95 | | 1 | | 1 | | 1 | |

Table 3. The experimental results of the approach

| Software and versions<br>Evaluation Measures | ArgoUML | | ArgoUML | | MALLET | | MALLET | |
|---|---|---|---|---|---|---|---|---|
| | 0.27.3 | 0.27.2 | 0.27.2 | 0.27.1 | 2.0.7 | 2.0.6 | 2.0.6 | 2.0.5 |
| Precision | 1 | | 0.996 | | 1 | | 0.992 | |
| Recall | 0.996 | | 0.993 | | 0.982 | | 0.992 | |

Take bluefish for example, mapping results of clone groups between bluefish 2.2.4 and bluefish 2.2.3 are shown in Figure 3.

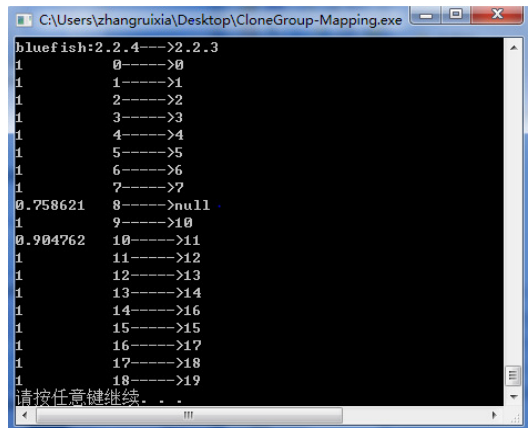

Figure 3. Mapping results of clone groups between Bluefish 2.2.4 and 2.2.3

The second and third column of the figure show clone group number of the corresponding version in Bluefish. There are 19(From 0 to 18) clone groups in Bluefish 2.2.4. There are 20(From 0 to 19) clone groups in Bluefish 2.2.3.The arrows indicate the corresponding clone group is traced to its origin clone group. For example, the 14th clone group of Bluefish 2.2.4 is mapped back to the 16th clone group of Bluefish 2.2.3.But the 14th clone group of Bluefish 2.2.4 is not traced to its origin clone group, which indicate that it is a new clone group, probably the great changes have taken place in its origin during the software evolution from Bluefish 2.2.3 to Bluefish 2.2.4, which similarity value between them is less than the threshold $\delta$. We note that 8th and 9th clone group of Bluefish2.2.3 do not appear in the list, probably they are removed or take great change during software evolution.

The first column of the figure show the largest similarity values of clone group topics between Bluefish 2.2.4 and Bluefish 2.2.3. If the value is not less than $\delta$ (0.8), There is a mapping relationship between them. It can be seen in the Figure 3 that most of the similarity value are as high as 1, namely most of the clone groups do not change during software evolution. Few of the similarity values are not 1, which indicate that clone codes have experienced some degree of change, such as the clone group is deleted, a few of clone fragments are added or removed.

## 6. DISCUSSION AND THREATS TO VALIDITY

**Limitations of clone detector.** The clone detector provides the basis data for clone group mapping, so clone group mapping approach directly is affected by clone detector. It is critical for clone group mapping to choose an accurate clone detector.

**Limitations of similarity threshold.** In the paper, similarity threshold between clone group topics across versions is determined based on the experience knowledge, and different software use the same similarity threshold, which have an impact on the results. Firstly, similarity threshold based on the experience knowledge can't reflect mapping efficiency of the algorithm. Secondly, the same threshold is used to different software that they exist remarkable differences in programming language, programming style and the degree of change between versions, which will reduce the validity of the mapping algorithm.

**The differences between versions.** It is discovered by the experimental results that the smaller differences between versions is, the higher accuracy the approach has. If clone group have happened so significant changes during software evolution that similarity value between two versions exceed the permitted threshold, which clone group that could have been traced to its origin clone group is failure to mapping. That is to say, Mutation or big difference between versions can reduce the accuracy of the mapping. Therefore, It contributes to improvement of accuracy of clone group mapping that using revision of software rather than release.

## 7. CONCLUSIONS

The activities of the programmers including copy, paste and modify result in lots of code clone in the software systems. However, Clone group mapping has a very important significance in the evolution of code clones. The clone group mapping approach based on topic modeling is proposed in the paper. By using topic modeling techniques to transform the mapping problem of high-dimensional code space into a low-dimensional topic space, the goal of clone group mapping was indirectly reached by mapping clone group topics. Experiments on 12 versions of 4 open source software show that the recall and precision are up to 0.99, thus the approach can effectively and accurately reach the goal of clone group mapping. The method is very effective for not only exact clone but also near-miss clones.

Ruixia Zhang, born in 1989, master, student at Inner Mongolia normal university. Her current research interests include software englneering, code analysis.

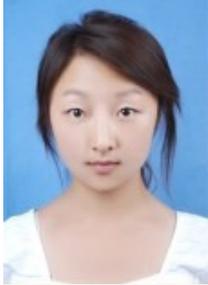

Huan Wang, born in 1991, master, student at Inner Mongolia normal university. His current research interests include software englneering, code analysis.

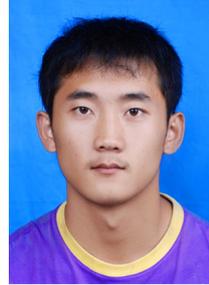

Liping Zhang, born in 1974, master, professor at Inner Mongolia normal university. Her current research interests include software englneering, code analysis.

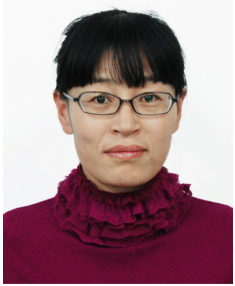

Zhuo Chen, born in 1989, master, student at Inner Mongolia normal university. His current research interests include software englneering, code analysis.

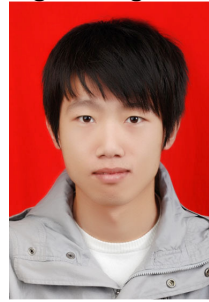